\documentclass{ifacconf}
\usepackage{graphicx}      
\usepackage{natbib}        

\usepackage{amsmath} 
\usepackage{amssymb}  

\DeclareMathAlphabet{\bit}{OML}{cmm}{b}{it}
\def\vect{\mathrm{vec}}           
\def\<{\leqslant}           
\def\>{\geqslant}           

\def\d{\partial}

\def\wt{\widetilde}

\def\Re{\mathrm{Re} }   
\def\Im{\mathrm{Im} }   

\def\cH{\mathcal{H}}   
\def\mA{{\mathbb A}}    
\def\mR{{\mathbb R}}    

\def\Tr{\mathrm{Tr}}       
\def\Sp{\mathrm{Sp}}       
\def\rT{{\rm T}}        

\def\bE{\mathbf{E}}    

\def\[[[{[\![\![}
\def\]]]{]\!]\!]}

\def\bra{\langle}
\def\ket{\rangle}

\def\re{{\rm e}}        
\def\rd{{\rm d}}        

\def\cL{\mathcal{L}}

\def\bJ{\mathbf{J}}

\def\x{\times}
\def\ox{\otimes}
\def\op{\oplus}

\def\fq{\mathfrak{q}}
\def\fp{\mathfrak{p}}
\def\fr{\mathfrak{r}}

\def\fB{\mathfrak{B}}

\def\fE{\mathfrak{E}}
\def\fF{\mathfrak{F}}

\def\fH{\mathfrak{H}}

\def\cZ{\mathcal{Z}}
\def\fZ{\mathfrak{Z}}

\def\sH{\mathsf{H}}
\def\sQ{\mathsf{Q}}

\def\bT{\mathbf{T}}

\def\cW{\mathcal{W}}

\def\cX{\mathcal{X}}

\def\cC{\mathcal{ C}}

\def\sP{\mathsf{P}}

\def\cP{\mathcal{P}}

\def\cA{\mathcal{ A}}
\def\cB{\mathcal{ B}}

\def\sA{\mathsf{A}}
\def\sB{\mathsf{B}}
\def\sC{\mathsf{C}}

\def\sX{\mathsf{X}}

\def\cS{{\mathcal S}}

\def\mS{{\mathbb S}}

\def\eps{\epsilon}
\def\ups{\upsilon}
\def\Ups{\Upsilon}

\begin{document}
\begin{frontmatter}

\title{
Coherent Quantum LQG Controllers with Luenberger Dynamics\thanksref{footnoteinfo}}

\thanks[footnoteinfo]{This work is supported by the Australian Research Council  grants  DP210101938, DP200102945.}

\author[First]{Igor G. Vladimirov${}^*$, \quad Ian R. Petersen}

\address[First]{School of Engineering, Australian National University, ACT 2601, Canberra, Australia    (e-mail: igor.g.vladimirov@gmail.com, i.r.petersen@gmail.com).}
\begin{abstract}
This paper is concerned with the coherent quantum linear-quadratic-Gaussian control problem of minimising an infinite-horizon  mean square cost for a measurement-free field-mediated interconnection of a quantum plant with a stabilising quantum controller. The plant and the controller are multimode open quantum harmonic oscillators, governed by linear quantum stochastic differential equations and  coupled to each other and the external multichannel bosonic fields in the vacuum state. We discuss an interplay between the quantum physical realizability conditions and the Luenberger structure associated with the classical separation principle. This leads to a quadratic constraint on the controller gain matrices, which is formulated in the framework of a swapping transformation for the conjugate positions and momenta in the canonical representation of the controller variables. For the class of coherent quantum controllers with the Luenberger dynamics, we obtain first-order necessary conditions of optimality in the form of algebraic equations, involving a matrix-valued Lagrange multiplier.
\vspace{-6.5mm}
\end{abstract}
\vspace{-3mm}
\begin{keyword}
Coherent quantum LQG control,
physical realizability,
separation principle,
Luenberger controller,
optimality conditions.
\end{keyword}
\end{frontmatter}

\vspace{-3mm}
\section{Introduction}
\vspace{-3mm}
Open quantum harmonic oscillators (OQHOs), described  by Hudson-Parthasarathy linear quantum stochastic differential equations (QSDEs) (\cite{HP_1984,P_1992}), are the closest quantum me\-chanical counterparts of classical linear stochastic systems. However, unlike classical random processes, the dy\-namic variables of an OQHO are noncommuting self-ad\-jo\-int operators on an  infinite-dimensional Hilbert space, organised similarly to the pairs of conjugate position and momentum operators (\cite{S_1994}). The QSDE, which governs the OQHO,   is driven by a quantum Wiener process with noncommuting components on a symmetric Fock space, thus modelling the interaction of the system with an external bosonic quantum field.
The energy exchange in this interaction and the self-energy of the OQHO (pertaining to its internal dynamics) are described in terms of the system-field coupling operators and the Hamiltonian, parameterised by coupling and energy matrices. Together with the canonical commutation relations (CCRs) for the system variables, this parameterisation leads to a specific structure of the state-space matrices of the linear QSDE,  so that they must satisfy  physical realisability (PR) conditions (\cite{JNP_2008})  in order to correspond to a quantum oscillator with CCR preservation.

The PR constraints are a significant obstacle to solving coherent quantum feedback control problems, where a given quantum plant is in a measurement-free field-mediated or direct (\cite{ZJ_2011a}) interconnection with a  quantum controller, which has to stabilise the closed-loop system and meet optimality or robust performance criteria.  One of such settings is the coherent quantum LQG (CQLQG) control problem (\cite{NJP_2009}) of minimising an infinite-horizon  mean square cost (for the plant variables and the controller output) over stabilising coherent quantum controllers, where both the plant and the controller are OQHOs (for example, with the same number of dynamic variables). Its classical counterpart (\cite{KS_1972}) admits a  separation principle, which decomposes the optimal LQG controller into a Kalman filter for updating the  conditional expectations of the plant variables,  conditioned on the observations,  and an actuator using the current plant state estimate,  along with a pair of independent algebraic Riccati equations.

However,  the CQLQG control problem does not lend itself to this particular combination of classical stochastic filtering and dynamic programming approaches because of the PR constraints mentioned above and the nature of quantum probability  (\cite{H_2001}). The latter describes the statistical properties of quantum processes in terms of density operators (or quantum states) on the underlying Hilbert space, which are more complicated than the scalar-valued classical probability measures and lead to the absence of classical joint distributions and conditional expectations for noncommutative quantum variables. Also, unlike classical observations, the noncommutative output fields of the quantum plant,  which drive the coherent quantum controller, are not accessible to simultaneous measurement. On the other hand, the absence of measurements (which are accompanied by  back-action effects and decoherence as   the loss of quantum information) is an advantage of coherent quantum control by interconnection compared to the classical observation-actuation paradigm using digital signal processing.


The motivation behind the CQLQG control problem  and the issue of obtaining an efficient  solution for it explain the recurrent research interest to this problem (and its feedback-free versions on coherent quantum filtering (\cite{MJ_2012,VP_2013b})) since its formulation in 2009. One of existing approaches to this problem is based on representing it as a constrained covariance control problem and applying variational methods of nonlinear functional analysis (in the form of Frechet differentiation of the mean square cost over the matrix-valued parameters (\cite{VP_2013a})) in combination with symplectic geometric and homotopy techniques to the development of optimality conditions and numerical algorithms (\cite{SVP_2017,VP_2021_MTNS}).
Although the CQLQG control  problem does not lend itself to a solution obeying the filtering-control separation principle with a Luenberger structure (\cite{L_1966}) (as a predictor-corrector scheme with a gain matrix with respect to an innovation process), the latter  was discussed  as an additional constraint, combined with the PR conditions,  for coherent quantum observers in (\cite{MJ_2012}).


The present paper extends these ideas to a class of coherent quantum controllers with Luenberger dynamics.
To this end, we use the freedom of assigning an arbitrary nonsingular CCR matrix to the controller variables (without affecting the LQG cost for the closed-loop system), including the negative of the CCR matrix of the plant variables. The latter is achieved by swapping the conjugate positions and momenta in the canonical representation of the quantum variables (or by applying the mirror reflections of (\cite{S_2000})). With the swapping transformation of the controller, the difference of the  plant and controller variables (which corresponds to the plant state estimation error in the case of classical optimal LQG controllers) forms a quantum process with zero one-point CCR matrix. Similarly to the classical case, this difference process conveniently replaces the plant variables in the closed-loop system for coherent quantum controllers of Luenberger type. The latter  imposes an additional constraint on the controller matrices in such a way that, together with the PR conditions,   the  gain matrices of the controller become dependent (through a quadratic constraint) and parameterise the dynamics and output matrices of the controller. This allows a matrix-valued Lagrange multiplier to be used in order to obtain first-order necessary conditions of optimality for this narrower class of coherent quantum controllers in the CQLQG control problem. The resulting optimality conditions involve a pair of coupled algebraic Lyapunov equations (ALEs) with block lower triangular matrices, which can simplify the analysis of their solution.


The paper is organised as follows.
Sec.~\ref{sec:plant_cont} specifies the class of quantum plants with field-mediated coherent quantum feedback.
Sec.~\ref{sec:PR} reviews the PR conditions and parameterization of the closed-loop system in terms of the energy and coupling matrices.
Sec.~\ref{sec:CQLQGinf} describes the CQLQG control problem.
Sec.~\ref{sec:swap} specifies the swapping transformation for the controller variables.
Sec.~\ref{sec:luen} discusses the class of coherent quantum controllers of Luenberger type.
Sec.~\ref{sec:lagrange} establishes first-order conditions of optimality for such controllers in the CQLQG control problem using the Lagrange multipliers.
Sec.~\ref{sec:conc} makes concluding remarks.

\section{Coherent Quantum Feedback}
\label{sec:plant_cont}
\begin{figure}[htbp]
\centering
\unitlength=1.25mm
\linethickness{0.2pt}
\begin{picture}(50.00,22.00)
    \put(7.5,8.5){\dashbox(35,13)[cc]{}}
    \put(10,11){\framebox(10,8)[cc]}
    \put(15,16){\makebox(0,0)[cc]{\small{quantum}}}
    \put(15,14){\makebox(0,0)[cc]{\small{plant}}}
    \put(35,16){\makebox(0,0)[cc]{\small{quantum}}}
    \put(35,14){\makebox(0,0)[cc]{\small{controller}}}
    \put(30,11){\framebox(10,8)[cc]}
    \put(0,15){\vector(1,0){10}}
    \put(50,15){\vector(-1,0){10}}
    \put(30,17){\vector(-1,0){10}}
    \put(20,13){\vector(1,0){10}}
    \put(-2,15){\makebox(0,0)[rc]{$w $}}
    \put(25,18){\makebox(0,0)[cb]{$\eta $}}
    \put(52,15){\makebox(0,0)[lc]{$\omega $}}
    \put(25,11.5){\makebox(0,0)[ct]{$y $}}
\end{picture}\vskip-12mm
\caption{A 
field-mediated interconnection of the quantum plant and coherent quantum controller,   interacting with each other (through their outputs $y$, $\eta$)  and with  
the quantum Wiener processes  $w $, $\omega $ which drive the QSDEs (\ref{x}), (\ref{y}), (\ref{xi}), (\ref{eta}). 
}
\label{fig:system}
\end{figure}
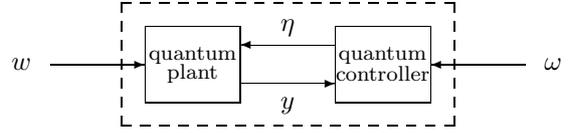

The CQLQG control setting (\cite{NJP_2009}) involves a quantum plant and a coherent quantum controller in the form of multimode OQHOs.  They are coupled to each other (see Fig.~\ref{fig:system})
through a measurement-free  feedback mediated by multichannel bosonic fields organised into column-vectors
\begin{equation}
\label{yeta}
    y
    :=
    (y_k)_{1\< k\< p_1},
    \qquad
    \eta
    :=
    (\eta_k)_{1\< k\< p_2}
\end{equation}
(the dependence on time $t$ is omitted for brevity)
and specified below. In this field-mediated interconnection, the plant and controller are also coupled to external bosonic fields modelled by self-adjoint
quantum Wiener processes  $w_1, \ldots, w_{m_1}$ and $\omega_1, \ldots, \omega_{m_2}$ (with even $m_1$,  $m_2$) on symmetric Fock spaces (\cite{HP_1984})  $\fF_1$, $\fF_2$, respectively. These quantum noises are assembled into vectors
\begin{equation}
\label{womega}
  w:=
  (w_k)_{1\< k \< m_1},
    \quad
  \omega:=
  (\omega_k)_{1\< k \< m_2},
    \quad
    \cW
    :=
    {\begin{bmatrix}
        w\\
        \omega
    \end{bmatrix}},
\end{equation}
where the augmented quantum Wiener process $\cW$ acts
on the composite Fock space $\fF:= \fF_1\ox \fF_2$ (with $\ox$ the tensor product of spaces or operators, including the Kronecker product of matrices),
and their future-pointing increments have the Ito tables
\begin{equation}
\label{wwww}
    \rd w\rd w^{\rT} = \Omega_1 \rd t,
    \
    \rd \omega\rd \omega^{\rT} = \Omega_2 \rd t,
    \
    \rd \cW \rd \cW^{\rT} = \Omega \rd t,
\end{equation}
where the transpose $(\cdot)^\rT$ applies to vectors or matrices of operators as if the latter were scalars.
Here, $\Omega_1$, $\Omega_2$, $\Omega$ are quantum Ito matrices given by
\begin{align}
\label{JJJ}
    \Omega_k
    & := I_{m_k} + iJ_k,
    \quad
    J_k:= I_{m_k/2} \ox \bJ,
    \quad
           \bJ
       :=
       {\begin{bmatrix}
         0 & 1\\
         -1 & 0
       \end{bmatrix}},
\\
\label{Om12}
    \Omega
    & :=
    {\begin{bmatrix}
        \Omega_1 & 0\\
        0 & \Omega_2
    \end{bmatrix}}
    =
    I_m + iJ,
    \quad
    J:=
    {\begin{bmatrix}
        J_1 & 0\\
        0 & J_2
    \end{bmatrix}}    ,
\end{align}
with     $m:= m_1+m_2$,
where $i:= \sqrt{-1}$ is the imaginary unit, and $I_r$ is the identity matrix of order $r$. 
The matrices $J_1 \in \mA_{m_1}$, $J_2\in \mA_{m_2}$, $J \in \mA_m$ in (\ref{JJJ}), (\ref{Om12}) (with $\mA_r$ the subspace of real antisymmetric matrices of order $r$) specify the CCRs
\begin{equation}
\label{wwcomm}
    [\rd w, \rd w^\rT]\!
     \! =\! 2iJ_1 \rd t,\,
    [\rd \omega, \rd \omega^\rT]\!
     \!=\!
    2iJ_2 \rd t,\,
    [\rd \cW, \rd \cW^\rT]\!
     \!=\! 2iJ \rd t,\!
\end{equation}
where $[\alpha, \beta^\rT] := ([\alpha_j,\beta_k])_{1\< j \< a, 1\< k \< b}$ is the matrix of commutators $[\alpha_j,\beta_k] = \alpha_j\beta_k - \beta_k\alpha_j$ between linear operators $\alpha_j$,  $\beta_k$ which form vectors $\alpha:= (\alpha_j)_{1\< j \< a}$, $\beta:= (\beta_k)_{1\< k\< b}$. The block diagonal structure of $J = \Im \Omega$ in (\ref{Om12}) comes from commutativity between the entries of $w$, $\omega$ acting on different Fock spaces.

The plant and the controller are endowed with initial Hilbert spaces $\fH_1$, $\fH_2$ and an even number $n$ of dynamic variables $x_1, \ldots, x_n$ and $\xi_1,\ldots, \xi_n$,  respectively, which are time-varying self-adjoint operators on the space
\begin{equation}
\label{fH}
    \fH:= \fH_0\ox \fF,
\end{equation}
where $\fH_0:= \fH_1 \ox \fH_2$ is the initial plant-controller space. With the same number $n$ of dynamic variables assumed for the plant and the controller (this plays an important role in what follows), $\frac{n}{2}$ counts their degrees of freedom.  The plant and controller variables are assembled into vectors
\begin{equation}
\label{xx}
    x
    :=
    (x_k)_{1\< k\< n},
    \qquad
    \xi
    :=
    (\xi_k)_{1\< k\< n},
    \qquad
    \cX
    :=
    {\begin{bmatrix}
        x\\
        \xi
    \end{bmatrix}}
\end{equation}
and satisfy the following CCRs with nonsingular matrices $\Theta_1, \Theta_2\in \mA_n$ and  $\Theta \in \mA_{2n}$:
\begin{equation}
\label{Theta12}
    [\cX,\cX^{\rT}]
    =
    {\begin{bmatrix}
      [x,x^{\rT}] & [x,\xi^{\rT}]\\
      [\xi,x^{\rT}] & [\xi,\xi^{\rT}]
    \end{bmatrix}}
    =
    2i\Theta,
    \quad
    \Theta
    :=
    {\begin{bmatrix}
      \Theta_1 & 0 \\
      0 & \Theta_2
    \end{bmatrix}}.
\end{equation}
In line with the block diagonal structure of $\Theta$, the plant variables commute with the controller variables (considered at the same moment of time): $[x,\xi^{\rT}]=0$,
since these operators  act initially (at time $t=0$)  on different spaces $\fH_1$, $\fH_2$, and the system-field evolution preserves the one-point CCRs. Accordingly,
the output fields $y_1, \ldots, y_{p_1}$ and $\eta_1, \ldots, \eta_{p_2}$ of the plant and the controller in (\ref{yeta}) are time-varying self-adjoint operators on the system-field space $\fH$ in (\ref{fH}). 
The Heisenberg dynamics of the internal and output variables of the plant are described by linear QSDEs
\begin{align}
\label{x}
    \rd x
    & =
    A x \rd t  +  B \rd w  + E \rd \eta ,\\
\label{y}
    \rd y
    & =
    C x \rd t  +  D \rd w ,
\end{align}
with given matrices
$A\in \mR^{n\x n}$,
    $B\in \mR^{n\x m_1}$,
    $C\in \mR^{p_1\x n}$,
    $D\in \mR^{p_1\x m_1}$,
    $E\in \mR^{n\x p_2}$.
The structure of $A$, $B$, $C$, $E$ will be specified in Sec.~\ref{sec:PR}.
The feedthrough matrix $D$ in (\ref{y}) is formed from conjugate pairs of rows of a permutation matrix of order $m_1$, so that $p_1$ is even and $p_1\< m_1$, with
\begin{equation}
\label{DDI}
    DD^{\rT} = I_{p_1}.
\end{equation}
The quantum Ito matrix $\wt{\Omega}_1$ of the plant output in (\ref{y}), defined by
$    \rd y \rd y^{\rT}
    =
    \wt{\Omega}_1 \rd t
$ (similarly to (\ref{wwww})),
is computed  in terms of (\ref{JJJ}) as     $\wt{\Omega}_1
    :=
    D\Omega_1 D^{\rT}
    = I_{p_1} + i\wt{J}_1$,  and its ima\-gi\-na\-ry part
\begin{equation}
\label{tJ1}
    \wt{J}_1:= DJ_1D^{\rT} = I_{p_1/2}\ox \bJ
\end{equation}
specifies the CCRs for the plant output $y$:
\begin{equation}
\label{ycomm}
    [\rd y, \rd y^\rT] = 2i\wt{J}_1 \rd t.
\end{equation}
The QSDE (\ref{x}) is driven by the external input field $w$ (as a quantum plant noise) and the controller output $\eta$, similar to the actuator signal  in  classical linear control (\cite{KS_1972}). The QSDE (\ref{y}) for the plant output $y$ resembles the equations for noise-corrupted observations with a ``signal'' part
\begin{equation}
\label{z}
    z
    :=
     Cx.
\end{equation}
However, the quantum process $y$ differs qualitatively from the classical observations  since the output  fields $y_1, \ldots, y_{p_1}$  are not accessible to simultaneous measurement as noncommuting quantum variables (\cite{H_2001}) in view of the relation
$    [y(s), y(t)^{\rT}] = 2i \min(s,t) \wt{J}_1$ for all
    $s,t\>0$, whose right-hand side vanishes only at $s=0$ or $t=0$.

The internal and output variables of the coherent quantum controller satisfy the linear QSDEs
\begin{align}
\label{xi}
    \rd \xi
     & =
    a\xi \rd t + b  \rd \omega  + e \rd y ,\\
\label{eta}
    \rd \eta
     & =
    c \xi  \rd t + d \rd \omega
\end{align}
(similar to the plant dynamics (\ref{x}), (\ref{y})),
with matrices
    $a  \in \mR^{n\x n }$,
    $b \in \mR^{n \x m_2}$,
    $c \in \mR^{p_2\x n }$,
    $d \in \mR^{p_2\x m_2}$,
    $e \in \mR^{n \x p_1}$,
where $b $, $e $ in (\ref{xi}) are the gain matrices of the controller with respect to the controller noise $\omega $ and the plant output $y $ in (\ref{y}).
Similarly to $D$ in (\ref{DDI}), the controller feedthrough matrix $d$ in (\ref{eta}) is also of full row rank and consists of conjugate pairs of rows of a permutation matrix of order $m_2$, so that $p_2$ is even and satisfies $p_2\< m_2$, along with
\begin{equation}
\label{ddI}
    dd^{\rT} = I_{p_2}.
\end{equation}
Accordingly, the quantum Ito matrix $\wt{\Omega}_2$ of the controller output fields in (\ref{eta}), defined by
$    \rd \eta \rd \eta^{\rT} = \wt{\Omega}_2 \rd t$
and computed as
$    \wt{\Omega}_2
    :=
    d\Omega_2 d^{\rT}
    = I_{p_2} + i\wt{J}_2$ in terms of (\ref{JJJ}), has the imaginary part
\begin{equation}
\label{tJ2}
    \wt{J}_2:= dJ_2d^{\rT} = I_{p_2/2}\ox \bJ,
\end{equation}
which, similarly to (\ref{ycomm}),  describes the CCRs for the cont\-roller output $\eta$:
\begin{equation}
\label{etacomm}
    [\rd \eta, \rd \eta^\rT] = 2i\wt{J}_2 \rd t.
\end{equation}
In what follows, the matrix $d$ (specifying the ``amount'' of noise $\omega$ in the controller output $\eta$) is fixed, while the matrices $a$, $b$, $c$, $e$ in (\ref{xi}), (\ref{eta}) can be varied subject to PR constraints of Sec.~\ref{sec:PR}. Similarly to (\ref{z}), the drift vector
\begin{equation}
\label{zeta}
    \zeta
     :=
    c \xi
\end{equation}
in (\ref{eta})
plays the role of a ``signal'' part of the controller output $\eta$ as a quantum noise-corrupted actuator process.

The QSDEs  (\ref{x}), (\ref{y}),  (\ref{xi}), (\ref{eta}) govern
the fully quantum closed-loop system in Fig.~\ref{fig:system}.
By analogy with classical LQG control,
the performance of the coherent quantum controller (with the process $\zeta $ in (\ref{zeta}) corresponding to the actuator signal) is described in Sec.~\ref{sec:CQLQGinf} in terms of a mean square cost functional for an auxiliary quantum process
\begin{equation}
\label{cZ}
    \cZ
     :=
     (\cZ_k)_{1\< k\< r}
     :=
    Fx   + G \zeta,
\end{equation}
where $
    F\in \mR^{r\x n}$,
$
    G\in \mR^{r\x p_2}
$ are given matrices. The entries of $\cZ$ are time-varying self-adjoint operators which are linear combinations of the plant variables and the controller output variables from (\ref{xx}), (\ref{zeta}) whose relative importance is specified by the weighting matrices
$
    F$,
$
    G
$. Similarly to the classical LQG control settings (\cite{KS_1972}),
the matrix $G$ is of full column rank:
\begin{equation}
\label{Grank}
    r\> \mathrm{rank} G = p_2,
\end{equation}
so that all  the entries of $\zeta$ are penalized through $G\zeta$ in (\ref{cZ}) for large mean square values.  The matrices $F$, $G$ are otherwise free from physical constraints, and their choice is part of the control design specifications.
The process $\cZ $ in  (\ref{cZ})  is expressed in terms of the  combined vector $\cX$ of the plant and controller variables in (\ref{xx}) and governed by
 \begin{equation}
\label{closed_ZX}
    \rd \cX
      =
      \cA      \cX \rd t +   \cB       \rd \cW ,
      \qquad
    \cZ
     =
      \cC        \cX ,
\end{equation}
where the QSDE is driven by the quantum Wiener process $\cW$ in (\ref{womega}) on the Fock space $\fF$. The matrices  $\cA\in \mR^{2n\x 2n }$, $\cB\in \mR^{2n \x m}$, $\cC\in \mR^{r\x 2n}$ of the closed-loop system (\ref{closed_ZX}) are obtained by combining the QSDEs (\ref{x}), (\ref{y}), (\ref{xi}), (\ref{eta}) with  (\ref{zeta}), (\ref{cZ}) as
\begin{equation}
\label{cABC}
    \cA
    :=
    {\begin{bmatrix}
        A & Ec\\
        eC & a
    \end{bmatrix}},
    \quad
    \cB
    :=
    {\begin{bmatrix}
        B & Ed\\
        eD & b
    \end{bmatrix}},
    \quad
    \cC
    :=
    {\begin{bmatrix}
        F & Gc
    \end{bmatrix}},
\end{equation}
similarly to the classical case.
While the matrices $F$, $G$ in (\ref{cZ}) can be arbitrary (subject to (\ref{Grank})),  the matrices $\cA$, $\cB$ of the QSDE in (\ref{closed_ZX}) are of specific structure which the fully quantum closed-loop system inherits from the plant and controller  (\cite{JNP_2008}), as reviewed in the next section.

\section{Physical Realizability Constraints}
\label{sec:PR}

The dynamics of the field-mediated coherent feedback interconnection are specified by the individual Hamiltonians
$\frac{1}{2} x^\rT R_1 x$,
$\frac{1}{2} \xi^\rT R_2 \xi$
and the vectors
${\scriptsize\begin{bmatrix}
  M_1\\ L_1
\end{bmatrix}} x$, ${\scriptsize\begin{bmatrix}
  M_2\\ L_2
\end{bmatrix}} \xi$ of operators of coupling of the plant and controller to the external fields and between each other. Here, $R_1\in \mS_n$ is the energy matrix of the plant (with $\mS_n$ the subspace of real symmetric matrices of order $n$), and   $M_1\in \mR^{m_1\x n}$, $L_1\in \mR^{p_2\x n}$ are the matrices of coupling of the plant with the external input field $w$ and the controller output $\eta$, respectively. Similarly, $R_2\in \mS_n $ is the energy matrix of the controller, and $M_2\in \mR^{m_2\x n }$, $L_2\in \mR^{p_1\x n }$ are the matrices of coupling of the controller with the external input field $\omega$ and the plant output $y$; see Fig.~\ref{fig:system}. These energy and coupling  matrices parameterise the plant matrices $A$, $B$, $C$, $E$ in (\ref{x}), (\ref{y}) and the controller matrices $a$, $b$, $c$, $e$ in (\ref{xi}), (\ref{eta}) as
\begin{align}
\label{AB}
    A
    =&
    2\Theta_1(R_1 + M_1^{\rT}J_1 M_1 + L_1^{\rT}\wt{J}_2L_1),
    \ \
    B   = 2\Theta_1 M_1^{\rT},\!\\
\label{CE}
    C =&2DJ_1 M_1,
    \quad
    E = 2\Theta_1 L_1^{\rT},\\
\label{ab}
    a
    =&
    2\Theta_2
    (R_2 + M_2^{\rT}J_2 M_2 + L_2^{\rT}\wt{J}_1L_2),
    \ \
    b  = 2\Theta_2 M_2^{\rT},\\
\label{ce}
    c = & 2dJ_2 M_2,
    \quad \ \,
    e  = 2\Theta_2 L_2^{\rT},
\end{align}
with the matrices $\wt{J}_1$, $\wt{J}_2$  given by (\ref{tJ1}), (\ref{tJ2}).
The special structure of the plant matrices in (\ref{AB}), (\ref{CE}) and the controller matrices in (\ref{ab}), (\ref{ce}) leads to the PR conditions for the plant:
\begin{align}
\label{PR1_plant}
    A \Theta_1 + \Theta_1 A^{\rT} + B J_1 B^{\rT} + E \wt{J}_2E^{\rT} &= 0,\\
\label{PR2_plant}
    C \Theta_1    + D J_1 B^{\rT}
    & =  0,
\end{align}
and similar conditions for the controller (\cite{JNP_2008}):
\begin{align}
\label{PR1_cont}
    a \Theta_2 + \Theta_2 a^{\rT} + b J_2 b^{\rT} + e \wt{J}_1 e^{\rT}   &= 0,\\
\label{PR2_cont}
    c \Theta_2 + d J_2 b^{\rT}& =  0,
\end{align}
with the PR constraints  (\ref{PR1_cont}), (\ref{PR2_cont})  on the controller mat\-rices $a$, $b$, $c$, $e$ (the matrix $d$ is fixed as mentioned before) being the distinctive feature of coherent quantum control formulations. 

The matrices $\cA$, $\cB$ of the closed-loop system (\ref{closed_ZX}), expres\-sed through the energy and coupling parameters by substituting (\ref{AB})--(\ref{ce}) into (\ref{cABC}), also satisfy PR conditions: 
\begin{equation}
\label{PR}
    \cA \Theta + \Theta \cA^{\rT}
    +
    \cB J \cB^{\rT}
    =0,
\end{equation}
which are similar to (\ref{PR1_plant}), (\ref{PR1_cont}) and secure the preservation of the CCRs (\ref{Theta12}).
Here, $J$ from (\ref{Om12}) is the CCR matrix for the combined quantum Wiener process $\cW$ in  (\ref{womega}). While the gain matrices $b$, $e$ of an arbitrary coherent quantum controller in (\ref{ab}), (\ref{ce}) (related by linear bijections to the coupling matrices $M_2$, $L_2$ since $\det\Theta_2\ne 0$)  are independent,
the matrices  $a$, $c$ of such a controller are parameterized by the triple $(R_2, b,e)\in \mS_n \x \mR^{n\x m_2}\x \mR^{n \x p_1}$ as
\begin{align}
\label{a_Rbe}
    a
     & =
    2\Theta_2 R_2
    -
    \frac{1}{2}
    (
        b J_2 b^{\rT}
        +
        e \wt{J}_1e^{\rT}
    )
    \Theta_2^{-1},\\
\label{c_b}
    c
    & =
    -d
    J_2
    b^{\rT}
    \Theta_2^{-1}
\end{align}
(see \cite{VP_2013a}).  
The relations (\ref{a_Rbe}), (\ref{c_b})
couple the matrices $a$, $c$ to $b$, $e$, thus making the stabilization  of the closed-loop system and the optimization of the coherent quantum controller (\ref{xi}), (\ref{eta}) qualitatively different from the classical control problems (irrespective of performance criteria). In particular, (\ref{c_b}) shows that an ``inflow'' of the  external quantum noise $\omega$ (through a nonzero gain matrix $b$) is essential in order for such a controller to produce a useful output $\eta$ with a nonzero drift vector $\zeta$ in (\ref{zeta}). At the same time, due to (\ref{ddI}) and the structure of $J_2 \in \mA_{m_2}$, $\Theta_2 \in \mA_n$ (satisfying $J_2^2 = -I_{m_2}$ and $\det \Theta_2\ne 0$) the linear map $\mR^{n \x m_2} \ni b\mapsto c \in \mR^{p_2\x n }$ in (\ref{c_b}) is surjective, so that any value of $c$ can be achieved by an appropriate choice of $b$ (for example, as $b = \Theta_2 c^\rT d J_2$).

The PR conditions (\ref{PR1_cont}), (\ref{PR2_cont}) impose const\-raints on the cont\-rol\-ler matrices $a$, $b$, $c$, $e$ even if the CCR matrix $\Theta_2 \in \mA_n$ is not specified. More precisely, if $a$ has no centrally symmetric  eigenvalues about the origin, 
and hence,  the Kronecker sum $a\op a := I_n\ox a + a\ox I_n$ is nonsingular, then $\Theta_2$ is recovered from (\ref{PR1_cont}) in terms of the vectorization $\vect(\Theta_2) = -(a\op a)^{-1}\vect(b J_2 b^{\rT} + e \wt{J}_1 e^{\rT})$,  and  its substitution into (\ref{c_b}) (assuming that $\det \Theta_2\ne 0$) makes the controller output matrix  $c$ a function of $a$, $b$, $e$.

\section{CQLQG control problem}
\label{sec:CQLQGinf}

Similarly to classical LQG control, 
the performance of the closed-loop quantum system (\ref{closed_ZX}) is described by the infinite-horizon mean square cost
\begin{equation}
\label{V}
    V
    \!:=\!
    \frac{1}{2}
    \lim_{T\to +\infty}
    \Big(\frac{1}{T}\int_0^T \bE (\cZ(t)^{\rT} \cZ(t)) \rd t\Big)
    \!=
    \frac{1}{2}\bra \cC^{\rT}\cC, \cP\ket\!\!\!
\end{equation}
(\cite{NJP_2009}),
where $\bra \cdot, \cdot\ket$ is the Frobenius inner product of matrices (\cite{HJ_2007}), and
\begin{equation}
\label{P}
    \cP:=
    \lim_{T\to +\infty}
    \Big(
        \frac{1}{T}
        \int_0^T
        \Re \bE (\cX(t)\cX(t)^\rT)
        \rd t
    \Big).
\end{equation}
The quantum expectation  $\bE \varphi := \Tr(\rho \varphi)$ is over the densi\-ty operator $\rho:= \rho_0\ox \ups$ on the system-field space $\fH$ in (\ref{fH}), where $\rho_0$ is the initial  plant-controller quantum state on $\fH_0$,  and $\ups$ is the vacuum field state on the Fock space $\fF$. 
The limits in (\ref{V}), (\ref{P}) exist whenever the initial plant and controller variables have finite second moments,
$\bE(\cX(0)^\rT\cX(0))
    <
    +\infty$,
and the closed-loop system is inter\-nally stable (the matrix $\cA$ in (\ref{cABC}) is
Hurwitz). In this case, $\cP$  is the controllability Gramian of the pair $(\cA, \cB)$: $\cP =       \int_0^{+\infty}
    \re^{t\cA} \cB\cB^{\rT}\re^{t\cA^{\rT}}
    \rd t$, 
found uniquely from the ALE
\begin{equation}
\label{cPALE}
  \cA \cP+ \cP\cA^\rT + \cB\cB^\rT = 0.
\end{equation}
Up to the factor of  $\frac{1}{2}$,  the cost $V$ in (\ref{V}) is the squared $\cH_2$-norm of a strictly proper  transfer function with the state-space realization triple $(\cA,\cB,\cC)$.
The CCR matrix $\Im \bE (\cX\cX^\rT) = \Theta$ from (\ref{Theta12}) does not contribute to (\ref{V}) since the subspaces $\mS_n$, $\mA_n$ in $\mR^{n\x n}$ are orthogonal  in the sense of $\bra\cdot, \cdot \ket$. The unique solution
$    \cS:= \cP + i\Theta
    =
    \int_0^{+\infty}
    \re^{t\cA}
    \cB\Omega \cB^{\rT}
    \re^{t\cA^{\rT}}
    \rd t
    \succcurlyeq 0
$ of the ALE $\cA \cS +\cS\cA^\rT   + \cB \Omega \cB^\rT = 0$ (which combines (\ref{PR}), (\ref{cPALE})),
is the quantum covariance matrix of the invariant zero-mean Gaussian state (\cite{KRP_2010})  for the closed-loop system variables.

The CQLQG control problem  (\cite{NJP_2009}) is formulated as the minimization
\begin{equation}
\label{CQLQG}
  V\to \inf
\end{equation}
of the cost (\ref{V})
over the controller matrices $a$, $b$, $c$, $e$ subject to the PR constraints (\ref{PR1_cont}), (\ref{PR2_cont}) and the internal stability condition that $\cA$ in (\ref{cABC}) is Hurwitz. Although the CCR matrix $\Theta_2$ of the controller variables in this problem is usually fixed, there is a certain freedom in its choice, which is exploited in what follows.

\section{Swapping in Controller Variables}
\label{sec:swap}

For any nonsingular matrix $\sigma \in \mR^{n \x n }$,  the transformation
\begin{equation}
\label{newxi}
    \xi\mapsto \sigma\xi,
    \qquad
    \Theta_2 \mapsto \sigma \Theta_2 \sigma^\rT
\end{equation}
of the controller variables and their CCR  matrix in (\ref{Theta12}),
with the energy and coupling matrices of the controller in (\ref{ab}), (\ref{ce})   being transformed as
    $R_2\mapsto \sigma^{-\rT}R_2\sigma^{-1}$,
    $M_2\mapsto M_2\sigma^{-1}$,
    $L_2\mapsto L_2\sigma^{-1}
$
(where $(\cdot)^{-\rT}:= ((\cdot)^{-1})^\rT$),
does not  affect the transfer function of the controller, and hence,  the cost $V$ in (\ref{V}) remains unchanged. Indeed, the matrices $\cC$, $\cP$ in (\ref{cABC}), (\ref{P}) are transformed by (\ref{newxi}) as
$
    \cC
    \mapsto
    \cC
    {\scriptsize\begin{bmatrix}
      I_n& 0\\
      0 & \sigma^{-1}
    \end{bmatrix}}$ and
    $
    \cP
    \mapsto
    {\scriptsize\begin{bmatrix}
      I_n& 0\\
      0 & \sigma
    \end{bmatrix}}
        \cP
    {\scriptsize\begin{bmatrix}
      I_n& 0\\
      0 & \sigma^\rT
    \end{bmatrix}}
$,
whereby $\cC \cP \cC^\rT$ remains the same and so also does $\bra \cC^\rT\cC, \cP\ket = \Tr(\cC \cP \cC^\rT)$ in (\ref{V}), thus implying the invariance of $V$. However, (\ref{newxi}) can be used in order to assign a given CCR matrix to the controller variables  (which is invariant only under the Lie group of symplectic similarity transformations identified  with the set $\Sp(\Theta_2)$ of matrices $\sigma$  satisfying
$\sigma \Theta_2 \sigma^\rT = \Theta_2$).  
Since the same also applies to the plant variables, there exist nonsingular matrices  $\sigma_1, \sigma_2 \in \mR^{n\x n}$  which convert the nonsingular CCR matrices $\Theta_1$, $\Theta_2$  to a canonical form:
\begin{equation}
\label{sig12}
    \sigma_1 \Theta_1 \sigma_1^\rT
    =
    \sigma_2 \Theta_2 \sigma_2^\rT
    =
    \frac{1}{2} I_{n /2} \ox \bJ
    =:
    \Ups,
\end{equation}
with $\bJ$ from (\ref{JJJ}).
The matrix $\Ups$ is the CCR matrix for $\frac{n }{2}$ con\-jugate position-momentum pairs $(\fq_k, \fp_k)$ (with com\-mutativity between them) assembled into a vector $\fr$ as
\begin{equation}
\label{pqr}
    \fr
    :=
    {\begin{bmatrix}
      \fr_1\\
      \vdots\\
      \fr_{n /2}
    \end{bmatrix}},
    \quad
    \fr_k
    :=
    {\begin{bmatrix}
      \fq_k \\
      \fp_k
    \end{bmatrix}},
\end{equation}
so that $[\fr,\fr^\rT] = 2i \Ups$, or equivalently,
$
    [\fr_j, \fr_k^\rT] = i\delta_{jk} \bJ$
for all
    $j, k = 1, \ldots, \frac{n }{2}
$,
where $\delta_{jk}$ is the Kronecker delta. By swapping the positions $\fq_k$ and momenta   $\fp_k$ in (\ref{pqr}), the vector $\fr$ is transformed  as
$
    \fr
    \mapsto
    \sigma_3 \fr$, with 
$
    \sigma_3
    :=
    I_{n/2}
    \ox
    {\scriptsize\begin{bmatrix}
        0 & 1\\
        1& 0
    \end{bmatrix}},
$
and acquires the CCR matrix $\sigma_3 \Ups \sigma_3^\rT = -\Ups$ in view of (\ref{sig12}). Therefore, the transformation matrix
$
    \sigma:= \sigma_1^{-1}\sigma_3\sigma_2
$
leads to
$
    \sigma \Theta_2 \sigma^\rT
     =
    \sigma_1^{-1}\sigma_3 \sigma_2
    \Theta_2
    \sigma_2^\rT \sigma_3^\rT \sigma_1^{-\rT}
     =
    -
    \sigma_1^{-1}
    \Ups
    \sigma_1^{-\rT}
    =
    -\Theta_1
$.
This transformation allows the controller variables $\xi_1, \ldots, \xi_n $ to be assumed for what follows  (without loss of generality,  except for the condition $\det \Theta_2\ne 0$) to have the CCR matrix
\begin{equation}
\label{TT}
    \Theta_2 = -\Theta_1.
\end{equation}
The same effect can be achieved by the mirror reflections $(\fq_k, \fp_k)\mapsto (\fq_k, -\fp_k)$ as in (\cite{S_2000}).
The relation (\ref{TT}) leads to commutativity between the entries (taken at the same moment of time)   of an auxiliary  quantum process
\begin{equation}
\label{eps}
  \eps
  :=
  x-\xi , 
\end{equation}
which corresponds to the plant state estimation error in the case of classical optimal LQG controllers  sa\-tis\-fying the separation principle (\cite{KS_1972}). More precisely, in view of (\ref{Theta12}), under the condition (\ref{TT}), the one-point CCR matrix of the difference process $\eps$ is zero:
\begin{align}
\nonumber
    [\eps,\eps^\rT]
    & =
    [x,x^\rT]
    -[x,\xi^\rT]
    -[\xi,x^\rT]
    +
    [\xi,\xi^\rT]\\
\label{epscomm}
    & =
    2i(\Theta_1+\Theta_2) = 0.
\end{align}
Nevertheless, $\eps$ is a substantially quantum process since $[x,\eps^\rT] = [x,x^\rT]-[x,\xi^\rT] = 2i\Theta_1\ne 0$ and also because (\ref{epscomm}) describes only the one-point CCRs for $\eps$, which does not prevent the two-point commutator matrix $[\eps(s),\eps(t)^\rT]$ from being nonzero at different moments of time $s\ne t $. The one-point CCRs for the processes $\eps$, $\xi$ take the form
\begin{equation}
\label{epsxicomm}
    [\sX,\sX^\rT]
  =
  2i
  \Xi,
  \qquad
  \Xi
  :=
  S\Theta S^\rT
    =
  {\begin{bmatrix}
    0 & \Theta_1\\
    \Theta_1 & - \Theta_1
  \end{bmatrix}},
\end{equation}
where
\begin{equation}
\label{SX}
    \sX
    :=
  {\begin{bmatrix}
    \eps\\
    \xi
  \end{bmatrix}}
    =
    S
    \cX,
    \qquad
    S:=
    {\begin{bmatrix}
      I_n & - I_n\\
      0 & I_n
    \end{bmatrix}}
\end{equation}
use the augmented vector $\cX$   of system variables from (\ref{xx}).

\section{Luenberger Type Controller Dynamics}
\label{sec:luen}

Consider a class of coherent quantum controllers of Luenberger type (\cite{L_1966}),   whose internal dynamics (\ref{xi}) is  represented as
\begin{equation}
\label{luen}
    \rd \xi
     =
    A\xi \rd t +  b  \rd \omega  + e (\rd y-C\xi\rd t) + Ec\xi\rd t ,
\end{equation}
in accordance with the plant dynamics (\ref{x}), (\ref{y}) and the structure of the controller  output (\ref{eta}). The Luenberger structure (\ref{luen}) imposes an additional constraint on the controller matrix $a$:
\begin{equation}
\label{a}
  a = A - eC + Ec.
\end{equation}
In this case, it is convenient to describe the closed-loop system dynamics in terms of the quantum processes $\eps$ from (\ref{eps}) and $\xi$. Since they are related to the vector $\cX$ in (\ref{xx}) by (\ref{SX}),
the matrix $\cA$ in (\ref{cABC}) is transformed to a block lower triangular form
\begin{align}
\nonumber
    \sA
    := &
    S \cA S^{-1}
    =
    {\begin{bmatrix}
      I_n & - I_n\\
      0 & I_n
    \end{bmatrix}}
    {\begin{bmatrix}
        A & Ec\\
        eC & a
    \end{bmatrix}}
    {\begin{bmatrix}
      I_n &  I_n\\
      0 & I_n
    \end{bmatrix}}\\
\label{cAluen}
     =&
    {\begin{bmatrix}
        A-eC & A-eC + Ec-a\\
        eC & eC+a
    \end{bmatrix}}=
    {\begin{bmatrix}
        A-eC & 0\\
        eC & A+Ec
    \end{bmatrix}},
\end{align}
where the last equality uses the Luenberger structure  (\ref{a}) of the matrix $a$.   The matrices $\cB$, $\cC$  in (\ref{cABC}) are transformed to
\begin{align}
\label{cBluen}
    \sB
    & :=
    S \cB
     =
    {\begin{bmatrix}
      I_n & - I_n\\
      0 & I_n
    \end{bmatrix}}
    {\begin{bmatrix}
        B & Ed\\
        eD & b
    \end{bmatrix}}
    =
    {\begin{bmatrix}
        B-eD & Ed-b\\
        eD & b
    \end{bmatrix}},\!\!\!\\
\label{cCluen}
    \sC
    & :=
    \cC S^{-1}
    =
    {\begin{bmatrix}
        F & Gc
    \end{bmatrix}}
    {\begin{bmatrix}
      I_n & I_n\\
      0 & I_n
    \end{bmatrix}}
    =
    {\begin{bmatrix}
        F & F+Gc
    \end{bmatrix}}.
\end{align}
The blocks of the matrices $\sA$, $\sB$, $\sC$  in (\ref{cAluen})--(\ref{cCluen}) describe the coefficients of the QSDEs for the processes $\eps$ in (\ref{eps}), $\xi$ in (\ref{luen}) and $\cZ$ in (\ref{closed_ZX}):
\begin{align}
\label{deps}
    \rd \eps
    & = (A-eC) \eps \rd t + (B-eD)\rd w + (Ed-b)\rd \omega, \\
\label{dxi}
    \rd \xi
     & =
    (eC\eps + (A+Ec)\xi) \rd t +  eD \rd w + b  \rd \omega,\\
\label{cZeps}
    \cZ
    & =
    F \eps + (F+Gc) \xi.
\end{align}
The QSDE (\ref{deps}) for the process $\eps$ is autonomous (does not involve $\xi$) since the matrix $\sA$ in (\ref{cAluen}) is block lower triangular. The latter makes the internal stability of the closed-loop system equivalent to the Hurwitz property of the matrices $A-eC$ and $A+Ec$,
as in the classical case. In order for these two conditions to be satisfied, it is necessary that the pair $(A,C)$ is detectable and $(A, E)$ is stabilizable. However, in the quantum case being considered,
\begin{equation}
\label{AEc}
    A+Ec = A+Ed J_2 b^\rT \Theta_1^{-1}
\end{equation}
is a function of $b$,  obtained by substituting (\ref{TT})  into (\ref{c_b}), with
\begin{equation}
\label{c_b1}
    c = dJ_2 b^\rT \Theta_1^{-1}.
\end{equation}
As a result, the fulfillment of the classical detectability  and stabilizability  conditions does not guarantee  the existence of controller gain matrices $b$, $e$ which make $A-eC$ and $A+Ec$ in (\ref{AEc}) Hurwitz,  since the Luenberger structure (\ref{a}),  combined with the PR conditions, leads to the following constraint  on $b$, $e$.

\begin{thm}
\label{th:beluen}
For the coherent quantum controller (\ref{xi}), (\ref{eta}) with the Luenberger structure (\ref{luen}), (\ref{a}) and the CCR matrix $\Theta_2$ in (\ref{TT}), the controller gain matrices $b$, $e$ are constrained by
\begin{equation}
\label{PRluen}
    (B-eD)J_1 (B-eD)^\rT + (Ed-b)J_2 (Ed-b)^\rT = 0.
\end{equation}
\end{thm}
\begin{pf}
From (\ref{TT}), (\ref{a}) and the second PR conditions (\ref{PR2_plant}), (\ref{PR2_cont}) for the plant and the controller, it follows that
\begin{align}
\nonumber
    a\Theta_2
    & =
    (A - eC + Ec)\Theta_2 = -A\Theta_1 + eC\Theta_1 + Ec\Theta_2\\
\label{aT2}
    & = -A\Theta_1 - eDJ_1 B^\rT - EdJ_2 b^\rT.
\end{align}
By using (\ref{aT2}) and the antisymmetry of the matrices $\Theta_1$, $J_1$, $J_2$ along with   the first PR condition (\ref{PR1_plant}) for the plant, the first PR condition (\ref{PR1_cont}) for the controller takes the form
\begin{align*}
    0  = &
    a\Theta_2 + \Theta_2 a^\rT + bJ_2 b^\rT + e\wt{J}_1 e^\rT\\
    = &
    -A\Theta_1 - eDJ_1 B^\rT - EdJ_2 b^\rT\\
    &
    -\Theta_1 A^\rT  - BJ_1 D^\rT e^\rT  - bJ_2 d^\rT E^\rT  + bJ_2 b^\rT + e\wt{J}_1 e^\rT    \\
    = &
    BJ_1 B^\rT + E\wt{J}_2 E^\rT
    - eDJ_1 B^\rT - EdJ_2 b^\rT\\
    &
    - BJ_1 D^\rT e^\rT  - bJ_2 d^\rT E^\rT  + bJ_2 b^\rT + e\wt{J}_1 e^\rT    \\
    =&
    (B-eD)J_1 (B-eD)^\rT + (Ed-b)J_2 (Ed-b)^\rT
\end{align*}
(with $\wt{J}_1$, $\wt{J}_2$ from (\ref{tJ1}), (\ref{tJ2})),
thus establishing (\ref{PRluen}).
  \hfill$\blacksquare$
\end{pf}

The relation (\ref{PRluen}) is equivalent to the preservation of the CCRs (\ref{epscomm}) for the process $\eps$ by the QSDE (\ref{deps}), which can also be seen from the first diagonal $(n\x n)$-block of the relation
$    \sA \Xi + \Xi \sA^{\rT}
    +
    \sB J \sB^{\rT}
    =0
$
obtained by representing (\ref{PR}) in terms of the matrices $\Xi$, $\sA$, $\sB$ from (\ref{epsxicomm}), (\ref{cAluen}), (\ref{cBluen}).

Now, by assembling the controller gain matrices $b$, $e$ into 
\begin{equation}
\label{be}
    \gamma
    :=
    \begin{bmatrix}
      b &  e
    \end{bmatrix}
    \in \mR^{n \x (m_2 + p_1)}
\end{equation}
and using $J$ from (\ref{Om12}),
the condition (\ref{PRluen}) is represented as
\begin{equation}
\label{PRluen1}
    f(\gamma)
    :=
    (\Gamma - \gamma \Delta)
    J
    (\Gamma - \gamma \Delta)^\rT    =0,
\end{equation}
where the matrices
\begin{equation}
\label{GD}
    \Gamma :=
    \begin{bmatrix}
      B & &Ed
    \end{bmatrix},
    \qquad
    \Delta:=
    {\begin{bmatrix}
      0 & I_{m_2}\\
       D & 0
    \end{bmatrix}}
\end{equation}
are associated with the plant gain and feedthrough matrices $B$, $E$, $D$ and the controller feedthrough matrix $d$ (which are fixed).
Completion of the square in (\ref{PRluen1}) yields
\begin{align}
\nonumber
    f(\gamma)
    & =
    \Gamma J \Gamma^\rT
        -
    \gamma \Delta J \Gamma^\rT
    -
    \Gamma J \Delta^\rT \gamma^\rT
    +
        \gamma K \gamma^\rT
    \\
\nonumber
    & =
    (\gamma - \gamma_0)K (\gamma - \gamma_0)^\rT + \mho\\
\label{PRluen2}
    & = (b-b_0)J_2 (b-b_0)^\rT+(e-e_0)\wt{J}_1 (e-e_0)^\rT + \mho,
\end{align}
where
\begin{align}
\label{gmho}
    \gamma_0
    & :=
    -\Gamma J \Delta^\rT K
    =
    \begin{bmatrix}
      b_0 &  e_0
    \end{bmatrix},
    \
    b_0 := Ed,
    \
    e_0 :=     -BJ_1D^\rT \wt{J}_1,\!\!\\
\label{mho}
    \mho
    & :=
    \Gamma(J + J\Delta^\rT K \Delta J)\Gamma^\rT,
\end{align}
and use is made of an orthogonal real antisymmetric mat\-rix
\begin{equation}
\label{K}
    K
    :=
    \Delta J \Delta^\rT
    =
    {\begin{bmatrix}
    J_2 & 0\\
    0 & \wt{J}_1
  \end{bmatrix}}
  =
  I_{(m_2+p_1)/2}\ox \bJ
\end{equation}
(so that $K^2 = -I_{m_2 + p_1}$), computed with the aid of (\ref{JJJ}), (\ref{Om12}), (\ref{tJ1}), (\ref{GD}). Similarly to (\ref{wwcomm}), (\ref{ycomm}), (\ref{etacomm}), the matrix $K$ specifies the joint CCRs for the controller noise $\omega$ and the plant output $y$ as
$
    \Big[
        {\scriptsize\begin{bmatrix}
          \rd\omega\\
          \rd y
        \end{bmatrix}},
        {\scriptsize\begin{bmatrix}
          \rd\omega\\
          \rd y
        \end{bmatrix}}^\rT
    \Big]
    =
    2iK \rd t
$.  In view of (\ref{PRluen1}), (\ref{PRluen2}), all the pairs $(b,e)$ satisfying (\ref{PRluen}) and organised  as in (\ref{be}) are described by the inclusion
\begin{equation}
\label{gall}
    \gamma \in \fZ(K,\mho)+\gamma_0,
\end{equation}
where, for any given matrix $\alpha \in \mA_n$,  the set
\begin{equation}
\label{fZ}
    \fZ(K,\alpha)
    :=
    \{\beta \in \mR^{n\x (m_2+p_1)}:\ \beta K\beta^{\rT} + \alpha=0\}
\end{equation}
is invariant under the right multiplication of its elements by symplectic matrices $\sigma\in \Sp(K)$ (whereby any $\beta \in \fZ(K,\alpha)$ is converted to $\beta \sigma \in \fZ(K,\alpha)$ since $\beta \sigma K  (\beta\sigma)^\rT = \beta K \beta^{\rT}=-\alpha$).

\begin{thm}
\label{th:beexist}
In addition to the assumptions of Theorem~\ref{th:beluen}, suppose the dimensions $m_2$, $p_1$ of the controller noise and the plant output   are large enough in the sense that
\begin{equation}
\label{nm2p1}
    m_2 + p_1 \> n.
\end{equation}
Then the controller gain matrices $b$, $e$  satisfying (\ref{PRluen}) exist and are described by (\ref{gall}) in terms of (\ref{be}), (\ref{GD}), (\ref{gmho})--(\ref{K}).
\end{thm}
\begin{pf}
By Lemma~\ref{lem:skew} of Appendix~\ref{sec:quadsol} applied to solvability of the equation $\beta K \beta^\rT = -\mho$ with $\mho \in \mA_n$ from (\ref{mho}) and the nonsingular matrix $K \in \mA_{m_2+p_1}$ in (\ref{K}), the condition (\ref{nm2p1}) implies that the set $\fZ(K,\mho)$ in (\ref{gall}) is nonempty. 
\hfill$\blacksquare$
\end{pf}

Since the set $\fZ(K,\mho)+\gamma_0$ (whose nonemptiness is guaranteed by (\ref{nm2p1})) is not an affine subspace, the existence of pairs $(b,e)$, which satisfy (\ref{gall}) and make the matrices $A-eC$ and $A+Ec$ in (\ref{AEc}) Hurwitz, is a nontrivial   open problem. In this regard, the following decompositions of the set (\ref{fZ})  can appear to be useful: \begin{align}
\nonumber
    \fZ(K,\alpha)
    & =
    \bigcup_{b \in \mR^{n\x m_2}} \{\begin{bmatrix}
      b & e
    \end{bmatrix}: e \in \fZ(\wt{J}_1, bJ_2 b^\rT + \alpha) \}\\
\label{fZsplit}
    & = \bigcup_{e \in \mR^{n\x p_1}} \{\begin{bmatrix}
      b & e
    \end{bmatrix}: b \in \fZ(J_2, e\wt{J}_1 e^\rT + \alpha) \},
\end{align}
which are obtained from (\ref{PRluen2}),
provided at least one of the conditions $m_2\> n$ or $p_1\> n$ holds (each of them is stronger than (\ref{nm2p1})). For example, if $p_1\> n$, application of Lemma~\ref{lem:skew} shows that the set $\fZ(\wt{J}_1, \beta J_2 \beta^\rT + \alpha)$ on the right-hand side of the first equality in (\ref{fZsplit}) is nonempty for any $\beta \in \mR^{n\x m_2}$. In this case,  the stabilization part of the CQLQG control problem in the class of coherent quantum controllers with Luenberger dynamics  is equivalent to finding a matrix $b \in \mR^{n\x m_2}$ such that the matrix $A+Ec$ in (\ref{AEc}) is Hurwitz (provided $(A,E)$ is stabilizable) and the nonempty set $\fZ(\wt{J}_1, (b-b_0)J_2 (b-b_0)^\rT + \mho) + e_0$  contains a matrix $e\in \mR^{n\x p_1}$ which makes $A-eC$ Hurwitz, provided $(A,C)$ is detectable.

\section{Necessary Conditions of Optimality}
\label{sec:lagrange}

For the class of coherent quantum controllers with Luenberger dynamics (\ref{luen}), (\ref{a}), the CQLQG control problem (\ref{CQLQG}) reduces to minimising the mean square cost $V$ in  (\ref{V}) over the controller gain matrices $b\in \mR^{n\x m_2}$, $e\in \mR^{n \x p_1}$ subject to the constraint (\ref{PRluen}) along with the internal stability condition that $A-eC$ and $A+Ec$ in (\ref{AEc}) are Hurwitz. The first-order necessary conditions of optimality for such controllers are those of stationarity for the Lagrange function $\mR^{n\x (m_2+p_1)}\x \mA_n \ni (\gamma,\lambda)\mapsto \cL \in \mR$ given  by
\begin{equation}
\label{cL}
    \cL
    := V
    +
    \frac{1}{2}
    \bra \lambda, f(\gamma)\ket,
\end{equation}
where the matrix $\gamma$ is defined by (\ref{be}), and  $\lambda$ is a Lagrange multiplier pertaining to the representation (\ref{PRluen1}) of the constraint (\ref{PRluen}) whose left-hand side is $\mA_n$-valued. The LQG cost $V$ in  (\ref{V}), which  is invariant under the transformation $\cX \mapsto \sX$ of the system variables in (\ref{SX}), can be computed for any stabilizing Luenberger controller as
\begin{equation}
\label{sV}
    V
    =
    \frac{1}{2}
    \bra \sC^{\rT}\sC, \sP\ket
    =
    \frac{1}{2}
    \bra \sB\sB^{\rT},  \sQ \ket
    =
    -\bra \sA, \sH\ket.
\end{equation}
Here, $\sP$, $\sQ$ are  the controllability and observability Gramians for
the matrix triple $(\sA,\sB,\sC)$ in (\ref{cAluen})--(\ref{cCluen}), satisfying the ALEs
\begin{equation}
\label{PQALE}
  \sA\sP+ \sP\sA^{\rT} + \sB\sB^{\rT} = 0,
  \qquad
  \sA^{\rT}\sQ+ \sQ\sA + \sC^{\rT}\sC = 0
\end{equation}
and giving rise to the  Hankelian
\begin{equation}
\label{sH}
  \sH := \sQ\sP,
\end{equation}
which is a diagonalizable matrix whose eigenvalues are the squared Hankel singular
values (\cite{KS_1972}). The matrices $\sP$, $\sQ$, $\sH$ (and related matrices)  are split into blocks $(\cdot)_{jk}$, block rows $(\cdot)_{j\bullet}$ and block co\-lumns $(\cdot)_{\bullet k}$, with $j,k=1,2$, in accordance with the partitioning of the matrices $\sA$, $\sB$, $\sC$ into blocks $\sA_{jk}$, $\sB_{jk}$, $\sC_k$ (for example, $\sA_{11} = A-eC$, $\sB_{21}=eD$ and $\sC_2 = F+Gc$).  The block lower triangular structure of the matrix $\sA$ in (\ref{cAluen}) (with $\sA_{12} = 0$) allows the ALEs (\ref{PQALE}) to be represented as
\begin{align}
\label{PALE11}
    \sA_{11}\sP_{11}  + \sP_{11}\sA_{11}^\rT +  \sB_{1\bullet}\sB_{1\bullet}^\rT = 0,\\
\label{PALE12}
    \sA_{11}\sP_{12}+  \sP_{11}\sA_{21}^\rT  + \sP_{12}\sA_{22}^\rT  +  \sB_{1\bullet}\sB_{2\bullet}^\rT = 0,\\
\label{PALE22}
    \sA_{21}\sP_{12}  + \sA_{22}\sP_{22} + \sP_{21}\sA_{21}^\rT +  \sP_{22}\sA_{22}^\rT + \sB_{2\bullet}\sB_{2\bullet}^\rT = 0,    \\
\label{QALE11}
    \sA_{11}^\rT\sQ_{11}  + \sA_{21}^\rT\sQ_{21} + \sQ_{11}\sA_{11}
    +
    \sQ_{12}\sA_{21} + \sC_1^\rT \sC_1 = 0,\\
\label{QALE21}
    \sA_{22}^\rT \sQ_{21}  + \sQ_{21}\sA_{11} +  \sQ_{22}\sA_{21} + \sC_2^\rT \sC_1 = 0,\\
\label{QALE22}
    \sA_{22}^\rT\sQ_{22}  + \sQ_{22}\sA_{22} +  \sC_2^\rT \sC_2 = 0.
\end{align}
For any stabilising Luenberger controller, the blocks $\sP_{11}, \sP_{12}=\sP_{21}^\rT, \sP_{22} \in \mR^{n\x n}$ of $\sP$  are computed by successively solving the ALE (\ref{PALE11}), the algebraic Sylvester equation (ASE) (\ref{PALE12}) and the ALE (\ref{PALE22}). In a similar fashion,  the blocks $\sQ_{22}, \sQ_{21}=\sQ_{12}^\rT, \sQ_{11}\in \mR^{n\x n}$   of $\sQ$ are obtained by solving the ALE (\ref{QALE22}), the ASE (\ref{QALE21}) and the ALE (\ref{QALE11}) and give rise to an auxiliary matrix
\begin{equation}
\label{q}
  q:= \sQ_{11} + \sQ_{22} - \sQ_{12}-\sQ_{21} =
  \begin{bmatrix}
    I_n & -I_n
  \end{bmatrix}
  \sQ
  \begin{bmatrix}
    I_n \\ -I_n
  \end{bmatrix}
  = q^\rT
  \succcurlyeq 0.
\end{equation}
 Associated with these matrices and the Lagrange multiplier $\lambda \in \mA_n$ from (\ref{cL}) are self-adjoint operators
\begin{align}
\label{fB}
  \fB
  & :=
  \[[[
    q, I_{m_2}
    \mid
    \Theta_1^{-1}\sP_{22}\Theta_1^{-1},
    J_2 d^\rT G^\rT G dJ_2
    \mid
    -\lambda, J_2
  \]]],\\
\label{fE}
  \fE
  & :=
  \[[[
    q, I_{p_1}
    \mid
    -\lambda, \wt{J}_1
  \]]]
\end{align}
on the Hilbert spaces $\mR^{n\x m_2}$, $\mR^{n\x p_1}$ (with the Frobenius inner product), respectively.   Here,
$\[[[\varphi_1, \psi_1 \mid \ldots \mid \varphi_s, \psi_s\]]]:= \sum_{k=1}^s \[[[\varphi_k, \psi_k\]]]$ is the sum of ``sandwich'' operators of the form
$\[[[\varphi, \psi\]]]$  specified by real matrices $\varphi$, $\psi$ and mapping an appropriately dimensioned real matrix $\vartheta$ to $\[[[\varphi, \psi\]]](\vartheta) := \varphi \vartheta \psi$ (so that  $ \[[[-\varphi, -\psi\]]]= \[[[\varphi, \psi\]]]$). The adjoint of such an operator is $\[[[\varphi, \psi\]]]^\dagger = \[[[\varphi^\rT, \psi^\rT\]]]$, and hence, $\[[[\varphi, \psi\]]]$ is self-adjoint whenever the matrices $\varphi$, $\psi$ are both symmetric or both antisymmetric   (see Section 7 and Appendix A of (\cite{VP_2013a})).

\begin{thm}
\label{th:stat}
Under the conditions of Theorem~\ref{th:beluen},
a stabilising coherent quantum controller with Luenberger dynamics (\ref{luen}), (\ref{a}) is a stationary point of the Lagrange function (\ref{cL}) for the CQLQG control problem (\ref{CQLQG}) if and only if it satisfies
\begin{align}
\nonumber
  (\sQ_{21} &-\sQ_{11})Ed + \lambda Ed J_2\\
\label{dLdb0}
  &  +
  \Theta_1^{-1} (\sH_{22}^\rT E + (\sP_{21}+\sP_{22})F^\rT G) d J_2 +
  \fB(b) = 0,\\
\nonumber
    (\sQ_{21}& -\sQ_{11})
    BD^\rT
    +\lambda BJ_1 D^\rT \\
\label{dLde0}
    & +   (\sH_{21}-\sH_{11}) C^\rT + \fE(e) = 0, \!\!\!\!\!
\end{align}
where the linear operators $\fB$, $\fE$ are  associated by (\ref{fB}), (\ref{fE})  with the Lagrange multiplier $\lambda \in \mA_n$ and the blocks of the Gramians $\sP$, $\sQ$ and the Hankelian $\sH$ in (\ref{PQALE})--(\ref{q}) for the closed-loop system (\ref{deps})--(\ref{cZeps}).
\end{thm}
\begin{pf}
The partial Frechet de\-rivatives of (\ref{sV}) over $\sA$, $\sB$, $\sC$ as independent variables are
\begin{equation}
\label{dVdABC}
    \d_\sA V = \sH,
    \qquad
    \d_\sB V = \sQ \sB,
    \qquad
    \d_\sC V = \sC \sP
\end{equation}
(see (\cite{SIG_1998})).
Similarly to (\cite{VP_2013a}), the chain rule differentiation of the cost (\ref{sV}) as a  composite function $b\mapsto (b,c)\mapsto (\sA, \sB, \sC)\mapsto V$  and $e\mapsto (\sA, \sB)\mapsto V$ of the independent variables $b$, $e$ using (\ref{cAluen})--(\ref{cCluen}),   (\ref{c_b1}), (\ref{dVdABC}) leads to
\begin{align}
\nonumber
  \d_b V
   = &
  (\d_b \sB)^\dagger(\d_\sB V)
  +
  (\d_bc)^\dagger((\d_c \sA)^\dagger(\d_\sA V)+(\d_c \sC)^\dagger(\d_\sC V))  \\
\nonumber
    = &
  (\sQ\sB)_{22}-(\sQ\sB)_{12}+
  \Theta_1^{-1}(E^\rT \sH_{22}+G^\rT \sC \sP_{\bullet 2})^\rT dJ_2 \\
\label{dbV}
  = &
  (\sQ_{2\bullet}-\sQ_{1\bullet})\sB_{\bullet 2} +
  \Theta_1^{-1} (\sH_{22}^\rT E + \sP_{2\bullet}\sC^\rT G) d J_2,\\
\nonumber
  \d_e V
  = &
  (\d_e \sA)^\dagger(\d_\sA V) +  (\d_e \sB)^\dagger(\d_\sB V)\\
\nonumber
    = &
  (\sH_{21}-\sH_{11}) C^\rT +
  ((\sQ\sB)_{21}-(\sQ\sB)_{11}) D^\rT\\
\label{deV}
  = &
  (\sH_{21}-\sH_{11}) C^\rT +
    (\sQ_{2\bullet}-\sQ_{1\bullet})
\sB_{\bullet 1}D^\rT.
\end{align}
Here, use is made of the identities $\sH_{jk} = \sQ_{j\bullet}\sP_{\bullet k}$ and  $(\sQ\sB)_{jk} = \sQ_{j\bullet}\sB_{\bullet k}$, along with 
the relation (\ref{c_b1}) represented in a sandwich operator form as $c = (\[[[dJ_2, \Theta_1^{-1}\]]]\circ \bT)(b)$, where $\bT(\cdot):= (\cdot)^\rT$ is the matrix transpose operator, so that $(\d_b c)^\dagger = \bT \circ \[[[J_2d^\rT , \Theta_1^{-1}\]]] = \[[[\Theta_1^{-1}, d J_2 \]]] \circ \bT$ in view of the antisymmetry of the matrices $\Theta_1$, $J_2$ and self-adjointness of $\bT$. By substituting (\ref{cBluen}), (\ref{cCluen}) into (\ref{dbV}), (\ref{deV}), it follows that
\begin{align}
\nonumber
  \d_b V
  = &
  (\sQ_{21}-\sQ_{11})Ed +
  \Theta_1^{-1} (\sH_{22}^\rT E \!+\! (\sP_{21}+\sP_{22})F^\rT G) d J_2\\
\label{dbV1}
  & +
    q b
    +
    \Theta_1^{-1}\sP_{22}\Theta_1^{-1}
    b
    J_2 d^\rT G^\rT G dJ_2,\\
\label{deV1}
  \d_e V
  = &
  (\sH_{21}-\sH_{11}) C^\rT +
    (\sQ_{21}-\sQ_{11})
    BD^\rT + qe.
\end{align}
where use is also made of the identities
$
    (\sQ_{2\bullet}-\sQ_{1\bullet})\sB_{\bullet 2} =
    {\small\begin{bmatrix}
      \sQ_{21}-\sQ_{11} & \sQ_{22}-\sQ_{12}
    \end{bmatrix}}
    {\scriptsize\begin{bmatrix}
        Ed-b\\
        b
    \end{bmatrix}}
    =
    (\sQ_{21}-\sQ_{11}) Ed +qb
$ and $
  (\sQ_{2\bullet}-\sQ_{1\bullet})
  \sB_{\bullet 1}D^\rT
    =
    (\sQ_{2\bullet}-\sQ_{1\bullet})
    {\scriptsize\begin{bmatrix}
        BD^\rT -e\\
        e
    \end{bmatrix}}
    =
    (\sQ_{21}-\sQ_{11}) BD^\rT  +qe
$
in view of (\ref{DDI}), (\ref{cBluen}), (\ref{q}). The Frechet differentiation of the constraint-related term of the Lagrange function $\cL$ in (\ref{cL}) with respect to  $\gamma$  in (\ref{be}) yields
\begin{align}
\nonumber
    \frac{1}{2}
    \d_\gamma
    \bra \lambda, f(\gamma)\ket
    & =
    \lambda (\gamma_0 - \gamma)K\\
\label{dconstdgamma}
    & =
    \lambda
    \begin{bmatrix}
      (Ed-b)J_2 &  &
      BJ_1 D^\rT - e\wt{J}_1
    \end{bmatrix},
\end{align}
where (\ref{GD})--(\ref{K}) are used. The corresponding partial Fre\-chet deri\-va\-tives in $b$, $e$ are recovered as the blocks of (\ref{dconstdgamma}):
\begin{align}
\label{dLdb}
    \frac{1}{2}
    \d_b
    \bra
    \lambda, f(\gamma)
    \ket
    & =
    \lambda Ed J_2 - \lambda b J_2,  \\
\label{dLde}
    \frac{1}{2}\d_e
    \bra
        \lambda,
        f(\gamma)
    \ket
    & =
    \lambda BJ_1 D^\rT - \lambda e  \wt{J}_1.
\end{align}
A combination of (\ref{dbV1}), (\ref{deV1}) with (\ref{dLdb}), (\ref{dLde}) and (\ref{fB}), (\ref{fE}) leads to
\begin{align}
\nonumber
  \d_b \cL
  = &
  (\sQ_{21}-\sQ_{11})Ed + \lambda Ed J_2\\
\label{dLdb1}
  &  +
  \Theta_1^{-1} (\sH_{22}^\rT E + (\sP_{21}+\sP_{22})F^\rT G) d J_2 +
  \fB(b),\!\\
\nonumber
  \d_e \cL
  = &
    (\sQ_{21}-\sQ_{11})
    BD^\rT
    +\lambda BJ_1 D^\rT\\
\label{dLde1}
    & +   (\sH_{21}-\sH_{11}) C^\rT + \fE(e).
\end{align}
The conditions of stationarity (\ref{dLdb0}), (\ref{dLde0}) are now obtained by equating the Frechet derivatives of the Lagrange function in (\ref{dLdb1}), (\ref{dLde1}) to zero.
\hfill$\blacksquare$
\end{pf}

The first-order necessary conditions of optimality for the CQLQG control problem in the class of Luenberger controllers, provided by Theorem~\ref{th:stat}, form a set of nonlinear algebraic equations for the controller gain matrices $b$, $e$ and the Lagrange multiplier $\lambda$. They include (\ref{PRluen}) and the ALEs (\ref{PQALE}) which are coupled through (\ref{dLdb0}), (\ref{dLde0}). These equations for a locally optimal coherent quantum controller can be solved numerically (for example, by using Newton or gradient descent iterative algorithms). Their theoretical analysis (as well as computational aspects) can benefit from the block triangular structure  of the ALEs for the Gramians in (\ref{PALE11})--(\ref{QALE22}) (which is a consequence of the Luenberger architecture) and will be discussed elsewhere.

\section{Conclusion}
\label{sec:conc}

In the context of the CQLQG control problem, we have considered
a swapping transformation for the controller variables, leading to a difference process with zero one-point CCR matrix. We have discussed the interplay between the quantum PR conditions and the classical Luenberger structure, resulting in an additional quadratic constraint on the controller gain matrices. For the class of coherent quantum controllers with Luenberger dynamics, we have obtained the first-order necessary conditions of optimality, 
which involve coupled ALEs along with a matrix-valued Lagrange multiplier and ``multisandwich'' operators on appropriate matrix spaces.


\appendix

\section{Special Quadratic Equations}
\label{sec:quadsol}    

The following lemma is used in the proof of Theorem~\ref{th:beexist} in Sec.~\ref{sec:luen}.

\begin{lem}
\label{lem:skew}
For any nonsingular matrix $K \in \mA_\mu$ and any matrix $\alpha\in \mA_\nu$ of even order $\nu\<\mu$, there exists a matrix $\beta\in \mR^{\nu\x \mu}$ satisfying
\begin{equation}
\label{alphaDelta}
    \beta K \beta^{\rT} = \alpha.
\end{equation}
\end{lem}
\begin{pf}
Since $K$ is a nonsingular real antisymmetric mat\-rix (and hence, its order $\mu$ is even), it is representable as
\begin{equation}
\label{Deltagamma}
    K = \psi (I_{\mu/2}\ox \bJ)\psi^{\rT}
\end{equation}
in terms of a nonsingular matrix $\psi \in \mR^{\mu\x \mu}$, with $\bJ$ from  (\ref{JJJ}). In a similar fashion, since $p$ is even and $\alpha\in \mA_p$, there exists a matrix $\varphi\in \mR^{\nu\x \nu}$ (singular if so is $\alpha$) such that
\begin{equation}
\label{alphavarphi}
  \alpha  = \varphi (I_{\nu/2}\ox \bJ)\varphi^{\rT}\\
   =
  \begin{bmatrix}
    \varphi & 0
  \end{bmatrix}
  (I_{\mu/2}\ox \bJ)
  {\begin{bmatrix}
    \varphi^{\rT}
    \\
    0
  \end{bmatrix}}.
\end{equation}
Due to the assumption $\nu\< \mu$, the last equality in (\ref{alphavarphi}) is obtained by padding $\varphi$ with zeros to a $(\nu\x \mu)$-matrix and  using the partitioning
$
    I_{\mu/2}\ox \bJ
    =
    {\scriptsize\begin{bmatrix}
        I_{\nu/2}\ox \bJ & 0 \\
        0 & I_{(\mu-\nu)/2}\ox \bJ
    \end{bmatrix}}
$.
Since $\det\psi\ne 0$, it follows from (\ref{Deltagamma}), (\ref{alphavarphi}) that (\ref{alphaDelta}) is satisfied, for example, with
$
    \beta :=
  \begin{bmatrix}
    \varphi & 0
  \end{bmatrix}
  \psi^{-1}
$.
\hfill$\blacksquare$
\end{pf}

A slight modification of the proof extends  Lemma~\ref{lem:skew} to the case when the order $\nu$ of the matrix $\alpha$ is odd.
\end{document}